\documentstyle[prc,aps,preprint]{revtex}

\draft
\begin{document}

\title{Berry's conjecture and information theory}
\author{C. Jarzynski}
\address{Theoretical Astrophysics, T-6, MS B288 \\
         Los Alamos National Laboratory \\
         Los Alamos, NM 87545 \\
         {\tt chrisj@t6-serv.lanl.gov}} 
\date{\today}

\maketitle

\begin{abstract}
It is shown that, by applying a principle of information 
theory, one obtains Berry's conjecture regarding the 
high-lying quantal energy eigenstates of classically 
chaotic systems.
\end{abstract}


In many problems of physical interest, it is 
necessary to abandon a search for the exact solution, 
and to turn instead to a statistical approach.
This involves mentally replacing {\it the} answer
which we seek, with an ensemble of possibilities,
then adopting the attitude that each member of
the ensemble is an equally likely candidate for the
true solution.
The choice of ensemble then becomes centrally 
important, and here {\it information theory} provides
a reliable guiding principle. 
The principle instructs us to choose the 
least biased ensemble (the one which 
minimizes information content), subject to some
relevant constraints.
A well-known illustration 
arises in classical statistical mechanics:
the least biased distribution in phase space,
subject to a fixed normalization and average
energy, is the canonical ensemble of 
Gibbs\cite{chandler}.
Another example appears in random matrix theory:
by minimizing the information content of an 
ensemble of matrices, subject to various simple
constraints, one obtains the standard random
matrix ensembles\cite{balian}.
The purpose of this paper is to point out that
{\it Berry's conjecture}\cite{conjecture} regarding 
the energy eigenstates of chaotic systems, also 
emerges naturally from this {\it principle of least bias}.

Berry's conjecture makes two assertions
regarding the high-lying energy eigenstates $\psi_E$
of quantal systems whose classical counterparts
are chaotic and ergodic\footnote{
By ``chaotic and ergodic'', we mean that 
all trajectories, except a set of measure zero,
chaotically and ergodically explore the surface
of constant energy in phase space.}: 
(1) Such eigenstates appear to be random
Gaussian functions $\psi({\bf x})$ 
on configuration space, 
(2) with two-point correlations given by
\begin{equation}
\label{eq:corrs}
\overline{
\psi^*\Biggl({\bf x}-{{\bf s}\over 2}\Biggr) 
\psi\Biggl({\bf x}+{{\bf s}\over 2}\Biggr)} 
= {1\over\Sigma} \int d{\bf p}
\,e^{i{\bf p}\cdot{\bf s}/\hbar}
\delta[E-H({\bf x},{\bf p})]
\end{equation}
Here, $E$ is the energy of the eigenstate,
$H({\bf x},{\bf p})$ is the classical Hamiltonian describing 
the system, and 
$\Sigma\equiv\int d{\bf x}\int d{\bf p}\,
\delta[E- H({\bf x},{\bf p})]$;
if the Hamiltonian is time-reversal-invariant,
then $\psi({\bf x})$ is a real random Gaussian
function, otherwise $\psi({\bf x})$ is a 
complex random Gaussian function.
Berry's conjecture thus uniquely specifies, for
a given energy $E$, an ensemble ${\cal M}_E$
of wavefunctions $\psi({\bf x})$
(i.e.\ ${\cal M}_E$ is the Gaussian ensemble with two-point
correlations given by Eq.\ref{eq:corrs}),
and states that an eigenstate $\psi_E$
at energy $E$ will look as if 
it were chosen randomly from this ensemble.

The correlations given by Eq.\ref{eq:corrs}
are motivated by considering the 
Wigner function\cite{wigner}
corresponding to the eigenstate $\psi_E$,
\begin{equation}
\label{eq:wigner}
W_E({\bf x},{\bf p}) \equiv
(2\pi\hbar)^{-D}\,\int d{\bf s}\,
\psi_E^*\Biggl({\bf x}-{{\bf s}\over 2}\Biggr) 
\psi_E\Biggl({\bf x}+{{\bf s}\over 2}\Biggr)\,
e^{-i{\bf p}\cdot{\bf s}/\hbar},
\end{equation}
where $D$ is the dimensionality of the system.
For high-lying states $\psi_E$, this Wigner
function, after local smoothing in the
${\bf x}$ variable,
is expected to converge to the microcanonical 
distribution in phase space\cite{conjecture,shnir,voros,lh81}:
\begin{equation}
\label{eq:micro}
W_E^{\rm sm}({\bf x},{\bf p}) \approx
{1\over\Sigma}\,\delta[E-H({\bf x},{\bf p})].
\end{equation}
By taking the Fourier transform of both sides of
Eq.\ref{eq:corrs}, and then smoothing locally in
the $\bf x$-variable\footnote
{This smoothing is performed on a scale
which is large compared with the local correlation
length of $\psi({\bf x})$, 
but small compared with a classically 
relevant distance scale
(see e.g.\ equations 3.29 and 3.30 
of Ref.\cite{lh81}).
This allows us to replace ensemble averaging
with local smoothing.}
rather than averaging over the ensemble ${\cal M}_E$,
it is straightforward to show that 
the correlations given by Eq.\ref{eq:corrs} produce 
the desired result, Eq.\ref{eq:micro}.

The assertion that $\psi_E({\bf x})$ is a Gaussian
random function is most easily motivated by viewing 
$\psi_E({\bf x})$, locally, as a superposition
of de Broglie waves with random phases\cite{conjecture}.
When the number of these waves
becomes infinite, the central limit theorem tells us
that $\psi_E({\bf x})$ will look like a Gaussian
random function.

We can interpret Berry's conjecture as making a 
specific prediction about the eigenstate $\psi_E$:
once we compute $\psi_E({\bf x})$ by solving the
Schr\" odinger equation, we can subject it to 
various tests (see e.g.\ Ref.\cite{sred_gau}),
and we will observe that,
yes, $\psi_E({\bf x})$ really behaves as a Gaussian
random function with two-point correlations given
by Eq.\ref{eq:corrs}.
Alternatively, we can interpret Berry's conjecture
as providing us with the appropriate ensemble of
wavefunctions from which to choose a surrogate for
the true eigenstate $\psi_E$, 
if we cannot (or do not care to) actually
solve for $\psi_E({\bf x})$. 
In this interpretation, ${\cal M}_E$ stands to
$\psi_E$ much as, in classical statistical mechanics,
the canonical ensemble stands to the instantaneous
microscopic state of a system at a given temperature.
It is within the context of the second point of view
that we will show that Berry's conjecture may be
``derived'' from information theory.
Specifically, we will show that, by applying the
principle of least bias, and accepting the correlations
given by Eq.\ref{eq:corrs} as a set of relevant 
constraints, we are led immediately to a 
statement of Berry's conjecture.

We thus pose the following question.
Suppose we have a quantal Hamiltonian $\hat H$,
whose classical counterpart $H({\bf x},{\bf p})$
is chaotic and ergodic;
and suppose we are told that a high-lying
eigenstate of $\hat H$ --- represented by a 
wave function $\psi_E({\bf x})$ --- exists at
energy $E$.
Given this limited knowledge, how to we go 
about making a ``best guess'' for $\psi_E({\bf x})$?
By a best guess, we mean not
a single wave function, but rather a probability
distribution $P_E[\psi]$ in Hilbert space,
such that, by sampling randomly from this 
distribution, we are making a guess which takes
into account our limited knowledge regarding
$\psi_E$, but is otherwise unbiased.
Information theory provides a general
prescription for constructing such a 
distribution.
First, we quantify the information $I$ contained 
in an arbitrary distribution $P[\psi]$.
Next, we identify the constraints on $P[\psi]$
imposed by our limited knowledge.
Finally, we minimize $I\{P[\psi]\}$
subject to these constraints.
The resulting distribution $P_E[\psi]$ is the least
biased one, consistent with our limited knowledge.
Let us now implement this procedure.

Given a probability distribution $P[\psi]$ 
in Hilbert space, 
the amount of information $I$ contained in 
this distribution is:
\begin{equation}
\label{eq:i}
I\{P[\psi]\}
= \int P[\psi]\,\ln P[\psi].
\end{equation}
The integral is over all square-integrable functions
$\psi({\bf x})$, where $\psi({\bf x})$ is taken to be 
real if the Hamiltonian is time-reversal-invariant,
and complex otherwise.
[The integral in Eq.\ref{eq:i} requires a
measure $d\mu$ on Hilbert space.
We take the usual Euclidean measure of
field theories\cite{measure}:
a wavefunction is represented by its value at $N$
discrete points in configuration space,
and the set of these values is regarded as a
(real or complex) Cartesian vector.
Hence, $d\mu = d\psi_1\,d\psi_2\cdots d\psi_N$,
where $\psi_i\equiv\psi({\bf x}_i)$.
The limit $N\rightarrow\infty$ is finally taken.]

Since we will want to minimize $I\{P\}$ subject 
to relevant constraints on the distribution $P[\psi]$, 
our next task is to identify those constraints. 
The first is simply that $P$ ought to be 
normalized to unity:
\begin{equation}
\label{eq:con1}
\int P[\psi] = 1.
\end{equation}
The second constraint is that embodied by 
Eq.\ref{eq:corrs}:
\begin{equation}
\label{eq:con2}
\int P[\psi]\,\,
\psi^*({\bf x}_1) 
\psi({\bf x}_2)
= {1\over\Sigma} \int d{\bf p}
\,e^{i{\bf p}\cdot{\bf s}/\hbar}
\delta[E-H({\bf x},{\bf p})],
\end{equation}
where ${\bf s}\equiv {\bf x}_2 - {\bf x}_1$ and
${\bf x}\equiv ({\bf x}_1+{\bf x}_2)/2$.
(Both ${\bf x}_1$ and ${\bf x}_2$ are assumed to
be within the classically allowed region;
outside this region, the wavefunction is taken
to be zero.)
As explained briefly above 
(see also Refs.\cite{conjecture,voros,lh81}),
this constraint is motivated by the
expectation that the smoothed Wigner function
corresponding to $\psi_E$ will approximate the 
microcanonical distribution in phase space
(Eq.\ref{eq:micro}).
Note that Eq.\ref{eq:con2} does not represent
a {\it single} constraint, but rather a 
set of constraints, where
each member of the set is specified by
$({\bf x}_1,{\bf x}_2)$.

Finally, we minimize the information 
$I\{P[\psi]\}$, subject to the constraints in
Eqs.\ref{eq:con1} and \ref{eq:con2}.
We do this in the usual way, by introducing
Lagrange multipliers.
That is, we define
\begin{eqnarray}
A\{P[\psi]\}&\equiv& I\{P[\psi]\}
+\lambda\int P[\psi]
+\int\!\!\int
\Lambda({\bf x}_1,{\bf x}_2) 
\int P[\psi]\,
\psi^*({\bf x_1})\psi({\bf x}_2)\nonumber\\
&=& \int 
P[\psi] \Biggl(
\ln P[\psi] + \lambda + \int\!\!\int
\Lambda({\bf x}_1,{\bf x}_2)\psi^*({\bf x}_1)\psi({\bf x}_2)
\Biggr),
\end{eqnarray}
where $\lambda$ is the Lagrange multiplier associated
with Eq.\ref{eq:con1}, $\Lambda({\bf x}_1,{\bf x}_2)$
is the set of multipliers associated
with Eq.\ref{eq:con2}, and $\int\!\!\int$ is shorthand for
$\int d{\bf x}_1\int d{\bf x}_2$.
For a given distribution $P[\psi]$, the change
in $A$ induced by a small variation $\delta P[\psi]$ is,
to first order in $\delta P[\psi]$, 
\begin{equation}
\label{eq:var}
\delta A = \int \delta P\,
\Biggl(\ln P + (\lambda+1) + 
\int\!\!\int
\Lambda({\bf x}_1,{\bf x}_2)\psi^*({\bf x}_1)\psi({\bf x}_2)
\Biggr).
\end{equation}
To minimize $A$ (i.e.\, to minimize $I$ subject
to the constraints imposed by Eqs.\ref{eq:con1} and
\ref{eq:con2}) we insist that $\delta A = 0$
for all variations $\delta P$.
From Eq.\ref{eq:var}, it follows that
the distribution $P_E[\psi]$ which accomplishes this
minimization has the form:
\begin{equation}
\label{eq:final}
P_E[\psi] = {\cal N}
\exp - \int\!\!\int
\Lambda({\bf x}_1,{\bf x}_2)\psi^*({\bf x}_1)\psi({\bf x}_2).
\end{equation}
The constant $\cal N$ is 
determined by normalization (Eq.\ref{eq:con1}),
whereas the two-point correlations (Eq.\ref{eq:con2})
uniquely determine 
$\Lambda({\bf x}_1,{\bf x}_2)$.
(Specifically, the kernel
$\Lambda({\bf x}_1,{\bf x}_2)$ is just the inverse of 
$\overline{\psi^*({\bf x}_1)\psi({\bf x}_2)}$, divided 
by 2 if $H$ is time-reversal-invariant\cite{sred_com}.)

Once ${\cal N}$ and $\Lambda({\bf x}_1,{\bf x}_2)$ are 
determined, Eq.\ref{eq:final} completely specifies a 
probability distribution $P_E$ on Hilbert space.
By randomly sampling from this distribution,
we generate a random function $\psi({\bf x})$,
with two-point correlations 
$\overline{\psi^*({\bf x}_1)\psi({\bf x}_2)}$
which (by construction) satisfy Eq.\ref{eq:corrs}.
But is a function sampled from $P_E$ a
{\it Gaussian} random function?
The answer is yes\cite{sred_gau,sred_com}.
For a random function $f({\bf x})$, let
${\cal P}_n(f_1,\cdots,f_n)$ denote the
joint probability distribution of finding that
$f({\bf x}_i)=f_i$, $i=1,\cdots,n$.
Then $f({\bf x})$ is Gaussian if ${\cal P}_n$
is a Gaussian in $(f_1,\cdots,f_n)$-space,
for any $({\bf x}_1,{\bf x}_2,\cdots,{\bf x}_n)$,
$n\ge 1$\cite{kubo}.
A function $\psi({\bf x})$ obtained by sampling
the probability distribution given by
Eq.\ref{eq:final} satisfies this condition.

We thus arrive at the following conclusion:
given the limited knowledge that an eigenstate
of $\hat H$ exists at energy $E$, 
the least biased guess for $\psi_E({\bf x})$
(by reasonable construction)
is a Gaussian random function, with two-point
correlations given by Eq.\ref{eq:corrs}.
This is just another way of 
stating Berry's conjecture.
(Instead of saying that $\psi_E$ ``looks like''
a Gaussian random function,
we say that a Gaussian random function is a
``best guess'' for $\psi_E$.)
In this sense, Berry's conjecture allows for a 
statistical description of eigenfunctions,
by providing us with the appropriate ensemble
${\cal M}_E$
in Hilbert space to use as a stand-in for the
true energy eigenstate $\psi_E$.
When the latter is unobtainable\footnote{
Note that the computational effort required to 
solve Schr\" odinger's equation grows exponentially
with the dimensionality $N$ of configuration space.}
calculations performed with ${\cal M}_E$ may
be tractable\cite{sred_therm},
just as the canonical ensemble of ordinary
statistical mechanics makes possible accurate computations
without demanding detailed knowledge of the
microscopic state of the system.

The notion that Berry's conjecture is gainfully viewed
as a statistical theory ---
in analogy with classical statistical mechanics, or
random matrix theory --- has been a guiding theme
of this paper.
As stressed in the opening paragraph, the first
order of business with such theories is to identify
the proper ensemble to use in place of an exact
description of the object of study (be it the
microscopic state of a many-body system, or a 
complicated Hermitian matrix, or a quantal eigenstate).
A common feature of statistical theories is that
this ensemble follows in a natural way from
the information theoretic principle of least bias.
The purpose of this paper has simply been to point
out that Berry's conjecture shares this feature.

\section*{ACKNOWLEDGMENTS}

The author wishes to acknowledge useful conversations
and correspondence with S.Jain, J.Morehead, and M.Srednicki.


\begin{references}

\bibitem{chandler} D.Chandler, {\it Introduction to
Modern Statistical Mechanics}, section 3.7
(Oxford University Press, New York, 1987).
Note that minimizing the information content of the 
ensemble is the same as maximizing its entropy.

\bibitem{balian} R.Balian, Il Nuovo Cimento
{\bf 57B}, 183 (1968).
The author would like to thank S.Jain for bringing
this reference to his attention.

\bibitem{conjecture} M.V.Berry, J.Phys.A {\bf 10},
2083 (1977).

\bibitem{wigner} E.Wigner, Phys.Rev. {\bf 40}, 749 (1932).

\bibitem{shnir} A.I.Shnirelman, Ups.Mat.Nauk 
{\bf 29}, 181 (1974).

\bibitem{voros} A.Voros, Ann.Inst.Henri Poincar\' e
A {\bf 24}, 31 (1976); {\bf 26}, 343 (1977);
in {\it Stochastic Behavior in Classical and 
Quantal Hamiltonian Systems}, ed.\ by G.Casati
and J.Ford (Springer-Verlag, Berlin, 1979).

\bibitem{lh81} M.V.Berry, in {\it Les Houches
XXXVI, Chaotic Behavior of Deterministic Systems},
G.Iooss, R.H.G.Helleman, and R.Stora, eds.
(North-Holland, Amsterdam, 1983).
 
\bibitem{sred_gau} M.Srednicki and F.Stiernelof, 
J.Phys.A {\bf 29}, 5817 (1996), chao-dyn/9603012.

\bibitem{measure} L.S.Brown, {\it Quantum Field
Theory}, Chapter 1 (Cambridge University Press,
1992).

\bibitem{sred_com} M.Srednicki, Phys.Rev.E {\bf 54},
954 (1996), cond-mat/9512115.
 
\bibitem{kubo} R.Kubo, M.Toda, N.Hashisume,
{\it Statistical Physics II: Nonequilibrium 
Statistical Mechanics}, section 1.4
(Springer-Verlag, Berlin, 1985).

\bibitem{sred_therm} See e.g.\ the use of
Berry's conjecture to derive {\it eigenstate
thermalization}.
M.Srednicki, Phys.Rev.E {\bf 50}, 888 (1994),
cond-mat/9403051.

\end{references}
\end{document}